\documentclass{optica-article}

\journal{opticajournal} 

\articletype{Research Article}

\usepackage{lineno}

\usepackage{amsmath,amsfonts}
\usepackage{graphicx}
\usepackage{glossaries}
\usepackage{mathtools}
\usepackage{hhline}
\usepackage{makecell}
\usepackage{textcomp}
\usepackage{multirow}

\usepackage{glossaries}
\newacronym{paa}{PAA}{point-ahead angle}
\newacronym{leo}{LEO}{low-earth orbit}
\newacronym{meo}{MEO}{medium-earth orbit}
\newacronym{geo}{GEO}{geostationary-earth orbit}
\newacronym{fso}{FSO}{free-space optical}
\newacronym{ogs}{OGS}{optical ground station}
\newacronym{ao}{AO}{adaptive optics}
\newacronym{mse}{MSE}{mean-square error}
\newacronym{iqr}{IQR}{interquartile range}
\newacronym{wfs}{WFS}{wavefront sensor}

\begin{document}

\title{Uplink Pre-Compensation for Ground-to-Space Optical Communications Using Downlink Tomographic Reconstruction}

\author{
Alex Frost,\authormark{1,*} 
Nicolas Maron,\authormark{1}
Shane Walsh, \authormark{1} 
Benjamin Dix-Matthews, \authormark{1}
Sascha Schediwy \authormark{1} and
Michael Hart \authormark{2} 
}

\address{
\authormark{1}International Centre for Radio Astronomy Research, The University of Western Australia, Perth, Western Australia, Australia\\
\authormark{2}Hart Scientific Consulting International, Tucson, Arizona, The United States of America}

\email{\authormark{*}alex.frost@research.uwa.edu.au}


\begin{abstract*} 

Uplink pre-compensation in ground-to-satellite optical links remains a problem, with an appropriate measurement of the wavefront error in the uplink path often not being available. We present a method that uses successive measurements of the readily accessible downlink beam to perform a tomographic reconstruction of the volume of atmosphere common to the downlink beam and the ground-station field-of-view. These measurements are done through the existing downlink wavefront sensor. From here, an estimate of the wavefront error along the uplink path can be obtained. We evaluate this method in simulation over representative atmospheric conditions and find good performance, especially for situations where the satellite point-ahead angle is many times greater than the atmosphere's isoplanatic angle. Compared to pre-compensation directly using the downlink phase, we find that this method estimates the uplink path with a residual mean-square wavefront error that is up to 8.6 times less. The hardware simplicity of this method makes it a promising solution for uplink pre-compensation implementations targeted at optical communications.

\end{abstract*}


\section{Introduction}

\Gls{fso} laser links will revolutionise the speed and security of wireless signal transfer, promoting major advancements in the field of communications \cite{carrasco-casadoSpaceOpticalLinks2020}. However, higher optical frequencies have an increased sensitivity to atmospheric turbulence compared to radio- and microwave-frequency counterparts \cite{strohbehnLaserBeamPropagation1978}. Propagation through atmospheric turbulence causes rapid fluctuations in the intensity and phase of the beam, leading to frequent signal dropouts. Astronomy-derived \gls{ao} systems are often used in \gls{fso} links to alleviate these effects through active phase compensation of the received beam \cite{chenExperimentalDemonstrationSinglemode2015}. Such systems continuously measure the wavefront of some reference signal and correct it to a required shape.

In an \gls{fso} link between an orbiting satellite and an \gls{ogs}, \gls{ao} compensation on the satellite-to-ground downlink path can be done at the ground station using the downlink laser beam directly. However, the uplink path poses significant complications. Most atmospheric turbulence is concentrated in the first 20 km above the Earth's surface \cite{tysonPrinciplesAdaptiveOptics2022}.
This means the uplink beam will interact with turbulence immediately and then propagate a further >500 km to the satellite receiver. The resulting large spatial scale intensity fluctuations will cause fades over the entire satellite aperture. To overcome this, the uplink beam can be pre-compensated at the \gls{ogs} with a dedicated \gls{ao} system such that propagation through the atmosphere removes the turbulence-induced wavefront perturbations at the satellite. An experimental demonstration of pre-compensation has been conducted over a horizontal link, showing significantly improved received power statistics at the receiver \cite{bradyValidationPrecompensationPointaheadangle2019}.

The choice of reference signal for uplink pre-compensation is critical and situation-dependent. All satellites in orbit have their uplink and downlink paths separated by their \gls{paa}, a result of the satellite's orbital speed and the finite speed of light. Due to their low \gls{paa}, \Gls{geo} satellites generally have strong correlation between the phase perturbations in the uplink and downlink paths in low to moderate turbulence conditions. In these cases, the downlink signal serves as an excellent pre-compensation reference signal \cite{osbornAdaptiveOpticsPrecompensated2021}. However, as the satellite's orbital speed increases from the perspective of the \gls{ogs}, or the atmosphere becomes more turbulent, eventually the \gls{paa} exceeds the atmosphere's isoplanatic angle, $\theta_0$, and the two paths are considered uncorrelated \cite{tysonPrinciplesAdaptiveOptics2022}. The terrestrial \gls{fso} link pre-compensation demonstration from Brady et al. \cite{bradyValidationPrecompensationPointaheadangle2019} confirms that using the downlink signal for pre-compensation becomes less beneficial as the ratio $\mathrm{PAA/\theta_0}$ increases. For $\mathrm{PAA/\theta_0 >2}$, even the tip/tilt of the paths becomes uncorrelated, causing use of the downlink signal to ultimately worsen the received power statistics. 

For \gls{leo} satellites, where the uplink and downlink paths will effectively always be uncorellated, a signal at the \gls{paa} is preferred. This could be achieved through a laser guide star; however the inherent tip/tilt indetermination leaves a major contributor to the overall phase perturbations unsensed \cite{f.rigautLaserGuideStar1992}. Alternatively, the atmospheric structure at the \gls{paa} can be estimated by exploiting the readily available downlink beam. Simulation work has been conducted by Logone et al. \cite{lognonePhaseEstimationPointahead2023} which uses the downlink phase and intensity measurements and the overhead $\mathrm{C_n^2}$ profile to estimate the atmospheric structure at the \gls{paa}. 

We propose a new method that uses successive measurements of the downlink beam to perform a tomographic reconstruction of the atmosphere. The method assumes an atmosphere stratified into several equivalent layers with discrete apparent wind vectors, influenced by the orbital speed of the satellite. The net downlink phase is decomposed onto each layer, from which the phase at the \gls{paa} can be estimated. We present simulation results of this method, showing the reconstruction accuracy of the uplink path for a variety of different ground-to-space \gls{fso} link configurations. We additionally explore the method's sensitivity to errors in its input parameters. 

\section{Method}

\begin{figure} [ht]
\includegraphics{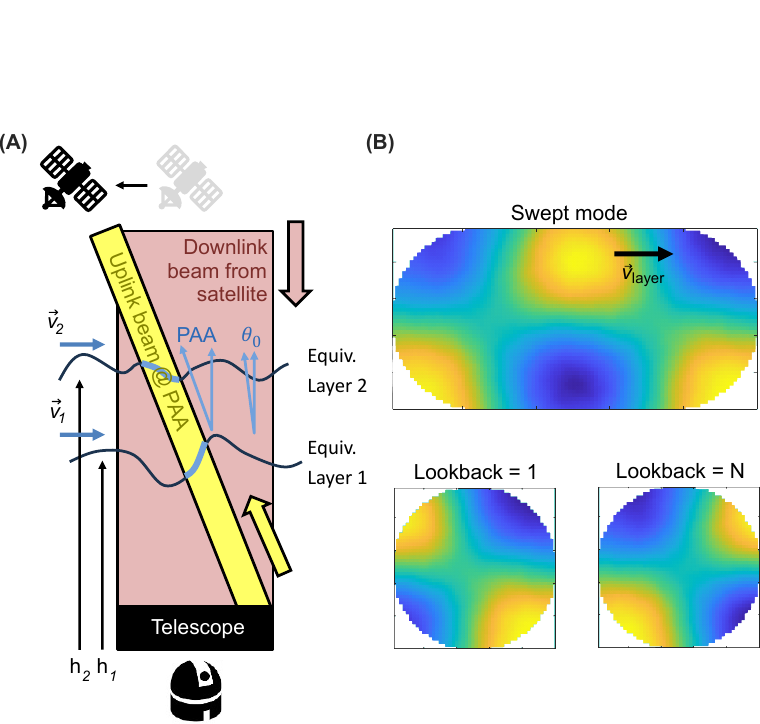}
\centering
\caption[] 
{ \label{fig:sys} 
(A) Schematic of this method. PAA -- point-ahead angle, $\theta_0$\,--\,isoplanatic angle, $\vec{v}$\,--\,equivalent layer velocity, $h$\,--\,equivalent layer height. (B) Above: visualisation of a single `swept mode' for a single equivalent layer, defined by the wind vector $\vec{v}_{\mathrm{layer}}$. Below: the sampling of the swept mode with the telescope aperture for lookback frames 1 and N.}
\end{figure}

A schematic of the method is shown in Fig.\;\ref{fig:sys}A. We assume an uplink beam transmitted from within a subaperture of the ground station's main aperture. The uplink beam is transmitted in such a way to maximise overlap with the downlink beam during propagation. This is because this method is only able to measure the volume of atmosphere above the telescope that has been illuminated by the downlink beam.

This method uses solely the apparent layer wind velocities (i.e., the combination of the actual layer wind and the pseudo-wind due to satellite tracking) to decompose the net downlink phase onto each equivalent layer. We assume a constant and homogeneous layer velocity over the measurement period, and as such are restricted to working within the frozen flow hypothesis timescale, $\tau_{fs}$, of $\sim$20\,ms \cite{schockMethodQuantitativeInvestigation2000}. This method allows us to map a series of consecutive \gls{wfs} measurements to one volumetric distribution of phase over the equivalent layers. We call these measurements `lookback frames'. 

The method is structured as follows: 
\begin{enumerate}
    \item  We define an area of atmosphere swept over the telescope aperture by the layer's wind vector over N lookback frames. This forms the envelope of the swept mode in Fig.\;\ref{fig:sys}B.
    \item Within this envelope, we generate an atmospheric phase perturbation basis by performing principal component analysis on 5000 realisations of Kolmogorov turbulence. We also explicitly add tip and tilt to this basis. 
    \item We sample across each mode of this basis with the telescope aperture over the N lookback frames. These samples give us information about how each mode evolves over the lookback frames.
    \item By measuring these samples with the \gls{wfs}, we build an influence matrix with dimensions $(\mathrm{2\times n_{subapp}\times N},\mathrm{m\times k})$, where each column maps all samples of a single mode of one equivalent layer to \gls{wfs} slopes. $\mathrm{n_{subapp}}$ is the number of \gls{wfs} subapertures; k is the number of equivalent layers; and m is the number of modes.
    \item We obtain the reconstructor matrix by computing the pseudoinverse of the influence matrix. The reconstructor matrix now allows us to decompose a series of downlink \gls{wfs} measurements onto each equivalent layer, thereby giving us a volumetric distribution of phase above the telescope.
    \item With the heights of each equivalent layer known, the estimate of the phase accrued along the uplink path is straightforward to compute.
\end{enumerate}

Importantly, the apparent wind velocities and equivalent layer heights can be obtained from the downlink \gls{wfs}, detailed in \cite{wangCharacterizingDaytimeWind2019} and \cite{hartAtmosphericTomographyArtificial2016}, respectively. This offers remarkable simplicity in a practical implementation of this method, with all necessary inputs to the method obtainable through the existing downlink \gls{ogs} hardware.

\section{Simulation Details}

To evaluate the performance of this method, we generate a sample space with fixed parameters listed in Table\,\ref{tab:fixed_params} and analyse the reconstruction accuracy of the phase along the uplink path across this sample space. We simulate at a wavelength of $1550$ nm, but provide equivalent $500$ nm values. 

We model a simple, two-layer atmosphere. Both layers have a Fried parameter of 100 mm. This results in a net Fried parameter of 66 mm, resembling that due to strong turbulence at Zenith in the Hufnagel-Valley turbulence profile (i.e., with $w = 21\, \mathrm{m/s},\ A = 1.7\times10^{-13}\;\mathrm{m^{-2/3}}$). We are careful to implement sub-harmonics in the phase screens as done by Lane and Dainty \cite{laneSimulationKolmogorovPhase1992} to obtain a \gls{mse} phase variance with significant tip/tilt contribution, given by {$\Delta\phi^2 = 1.03 (D/r_0)^{5/3}$} \cite{nollZernikePolynomialsAtmospheric1976}. 

\begin{table}[ht]
    \centering
    \caption{Fixed simulation parameters.}
    \begin{tabular}{|c|c|c|}
        \hline
        Parameter (units) & Symbol & Value at 1550 nm (500 nm)\\
        \hhline{|=|=|=|}
        Number of equivalent layers & $n_\mathrm{layer}$ & 2   \\
        Fried parameters (mm) & $r_{0,1},\;r_{0,2}$ & 100 (26), 100 (26) \\ 
        Layer heights (m) & $h_1,\;h_2$ & 2500, 5000 \\ 
        \hline
        Net Fried parameter (mm) & $r_0$ & 66 (17) \\
        Isoplanatic angle (") & $\theta_0$ & 1.1(0.28) \\
        \hline
        Downlink aperture diameter (mm) & $D_\mathrm{telescope}$ & 700 \\
        Uplink aperture diameter (mm) & $D_\mathrm{uplink}$ & 350 \\
        Wavefront sensor subapertures & $n_\mathrm{subapp}$ & 10$\times$10 \\
        Wavefront sensor framerate (kHz) & $f_\mathrm{WFS}$ & 1 \\ 
        \hline
    \end{tabular}
    
    \label{tab:fixed_params}
\end{table}

We choose layer heights of 2.5 km and 5.0 km, resulting in an isoplanatic angle of 1.1". This value is notably small, approximately five times less than that at Zenith for the Hufnagel-Valley strong turbulence profile. This value is chosen for computational efficiency, allowing large ratios of $\mathrm{PAA/\theta_0}$ to be tested without having to generate excessively large phase screens for the equivalent layers. 

We choose an \gls{ogs} configuration with a 700 mm aperture diameter and presently neglect any central obscuration. The \gls{wfs} is a  $10\times10$ subaperture Shack-Hartmann, chosen such that $D/r_0 \approx1$ for each subaperture, as is expected for a properly designed downlink \gls{ao} system.

Alongside these fixed parameters, we vary the equivalent layer apparent wind velocities and satellite \gls{paa} magnitude according to Table \ref{tab:variable_params}. The apparent wind speeds, dominated by the pseudo-wind due to satellite tracking, span $v_2/v_1$ ratios near what would be expected (i.e., $h_2/h_1$) for the layer heights simulated. The wind speeds are purposefully decoupled from the layer heights so that the effect of both variables on the uplink reconstruction can be analysed separately. The magnitudes of the apparent layer wind speeds roughly correspond to a satellite with a 5" \gls{paa}. While a 10" \gls{paa} is more representative of a \gls{leo} pass, simulating these wind speeds was computationally expensive due to the number of evolutions required for each phase screen. We have not observed a decrease in performance for high wind speeds, and so do not think this choice biases the analysis.

\begin{table}[ht!]
    \centering
    \caption{Variable simulation parameters.}
    \begin{tabular}{|c|c|c|}
        \hline
        Parameter (units) & Symbol & Values\\
        \hhline{|=|=|=|}
        \makecell{Equivalent layer apparent\\wind speeds (m/s)} & ($v_1$,\;$v_2$) & \makecell{(20, 22),\;(20, 30)\\(20, 45),\;(20, 60)}   \\
        \hline
        \makecell{Equivalent layer apparent wind\\direction relative to PAA (\textdegree)} & ($\theta_1$,\;$\theta_2$) & \makecell{(0, 0),\;(-15, 15)\\(-30, 30)} \\ 
        \hline
        Point-ahead angle (") & PAA & \makecell{2.2,\;5.4,\;10.8} \\
        \hline
        Lookback frames & N & \makecell{2,\;5,\;10,\;20} \\
        \hline
    \end{tabular}
    
    \label{tab:variable_params}
\end{table}

Additionally, we vary the angle between the equivalent layer wind vectors and the satellite's \gls{paa} for completeness, but note that the apparent wind direction per layer at these heights will be dominated by the psuedo-wind and will therefore be predominately in the \gls{paa} direction. The \gls{paa} values span typical values of \gls{leo} to \gls{geo} satellites \cite{mengaliGroundtoGEOOpticalFeeder2020, lazzaroEvaluationSatellitesPointAhead2022}, and the corresponding ratios of ${PAA/\theta_0}$ are also representative \cite{bonnefoisFeasibilityDemonstrationAO2022, eatonIsoplanaticAngleDirect1985}.

The number of lookback frames are chosen to be within the range of $\tau_{fs}$ given the 1 kHz frame-rate of the \gls{wfs}. Since non-frozen flow phase screen evolution is not simulated, this is critical in ensuring a fair evaluation of this method. 

We simulate 500 time steps per unique configuration of variables in Table \ref{tab:variable_params}, resulting in 81 separate simulations. Critically, we assume that any unsensed phase in the uplink path (i.e., that which lies outside of the downlink beam) is minimal. Considering diffraction-limited divergence of the uplink beam for this \gls{ogs} configuration, and a 10" \gls{paa}, the height range where the uplink beam lies partially outside the downlink footprint is 6.9 -- 15.5 km, resulting in an unsensed phase \gls{mse} of between 0.25 -- $0.02\;\mathrm{rad^2}$. As will be later shown in Fig.\;\ref{fig:histogram}, this is comparable to, or less than, the residual phase error due to perfect uplink pre-compensation, and is thus considered minimal.

\section{Results}

To evaluate the performance of this method, we compare the estimated phase along the uplink path to the actual simulated phase. We calculate the residual phase variance over the uplink aperture per simulation time step and build up a distribution. This is compared to three other pre-compensation cases, labelled in brackets in subsequent figures: 

\begin{itemize}
    \item Using the downlink beam directly (DL). This is considered the naive case as it does not account for any anisoplanatism between the downlink and uplink paths. As the downlink beam is already available, this case is practically simple to implement.
    \item Using the exact simulated phase along the uplink path (UL). This is the ideal case. Measuring this signal would require a beam transmitted from orbit at the \gls{paa}, or a laser guide star with tip/tilt information, and is therefore practically difficult to implement. 
    \item Applying no correction (NC). This is an important anchor to highlight cases where DL pre-compensation introduces additional error.
\end{itemize}

\begin{figure}[ht]
\includegraphics{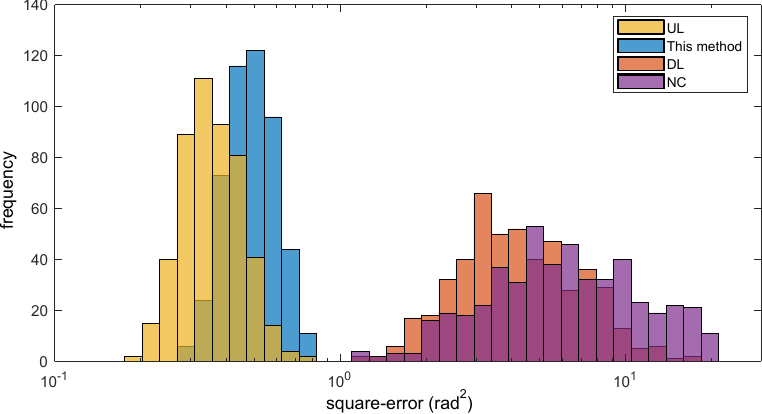}
\centering
\caption[] 
{\label{fig:histogram} 
Residual phase variance histograms for 500 time steps of a single simulation where $\mathrm{\theta_{layers}}$ = (0,0)\textdegree,\;$v_{\mathrm{layers}} =(20, 45)\ m/s$, $\mathrm{N = 10}$, \gls{paa} = 5.4" (PAA/$\theta_0$\,=\,5). Four histograms are shown for the four pre-compensation cases: this method (blue), DL (orange), UL (yellow), and NC (purple).}
\end{figure}

Any pre-compensation method that does not directly use the DL signal requires an additional deformable mirror. As such, the reduction in residual error with this method must be substantial to justify this extra cost. Distributions of the residual phase variances for all four cases over one simulation are shown in Fig.\;\ref{fig:histogram}. Here, we see a case where this method considerably outperforms the DL case, with a clear reduction in mean value and spread. Note that all signals are measured through the $10\times10$ subaperture \gls{wfs}, meaning that even pre-compensation in the UL case still has high spatial-frequency residual error.

Further analysis of this method is split into two parts. Firstly, the reconstruction accuracy of the uplink path is evaluated across the \gls{fso} link configurations from Table \ref{tab:variable_params}. Next, the robustness of this method to input parameter error is evaluated for one representative configuration.

\subsection{Uplink reconstruction over \gls{fso} configuration sample space}

\begin{figure}[ht]
\includegraphics{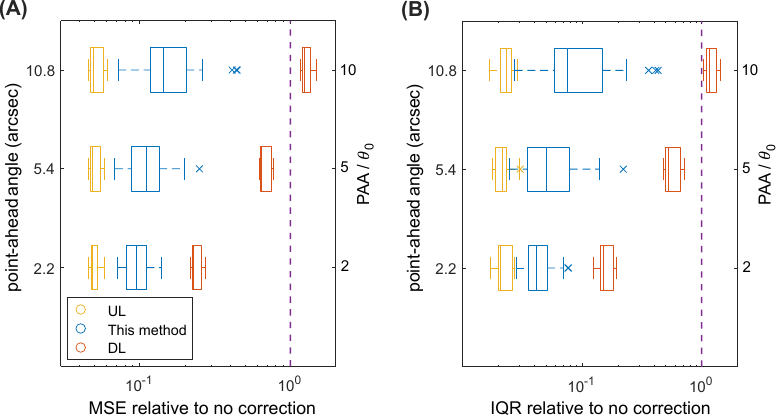}
\centering
\caption[] 
{ \label{fig:mse_iqr} 
Box and whisker plots for the (A) \Gls{mse} and (B) \gls{iqr} of the uplink phase reconstructions relative to the uncorrected uplink phase across all \gls{fso} link configurations in Tables \ref{tab:fixed_params} and \ref{tab:variable_params}. The dashed purple line at \gls{mse}, \gls{iqr} = 1 corresponds to the uncorrected uplink phase reference. Reconstructions using the three cases are shown: UL (yellow), this method (blue), and DL (orange).}
\end{figure}

The \acrfull{mse} and \gls{iqr} of the residual phase variances are calculated for every simulation configuration in Table \ref{tab:variable_params} and used to create Figs.\,\ref{fig:mse_iqr}A and B, respectively. Here, we plot the ratio of the \gls{mse} and \gls{iqr} for our three cases relative to the values with no correction applied. The ratios are collated into box and whisker plots, showing the spread of the reconstruction errors across simulation configurations. The ratios are grouped by the \gls{paa}. 

We see that this method has a lower \gls{mse} and \gls{iqr} than the DL case across all simulation configurations. As expected, this method does not outperform the UL case, but the reconstruction errors consistently remain closer to UL than to DL pre-compensation. The spread in performance across the simulated sample space is larger than for both the DL and UL cases. This is due to this method's dependence on equivalent layer velocities and positions. Changing link conditions will affect the reconstruction accuracy of the uplink path.

Looking at the DL case performance, we see that for $\mathrm{PAA/\theta_0 = 2}$ there is a significant reduction in \gls{mse} and \gls{iqr} with this naive approach, attributed to the downlink and uplink path being within the isokinetic angle \cite{sasielaElectromagenticWavePropagation2007}. For $\mathrm{PAA/\theta_0 = 5,10}$, the DL case either offers minimal improvement or results in worse performance, respectively, whereas this method consistently improves the reconstruction error.

Overall, these results demonstrate substantial improvement in the uplink path reconstruction error using this method compared to the DL case. Even for $\mathrm{PAA/\theta_0 =2}$, this method still reduces the \gls{iqr} by a factor of four. For increasing $\mathrm{PAA/\theta_0}$, this method performs increasingly better than DL pre-compensation, but strays further from the best-case performance of UL pre-compensation.

\begin{figure}[ht]
\includegraphics{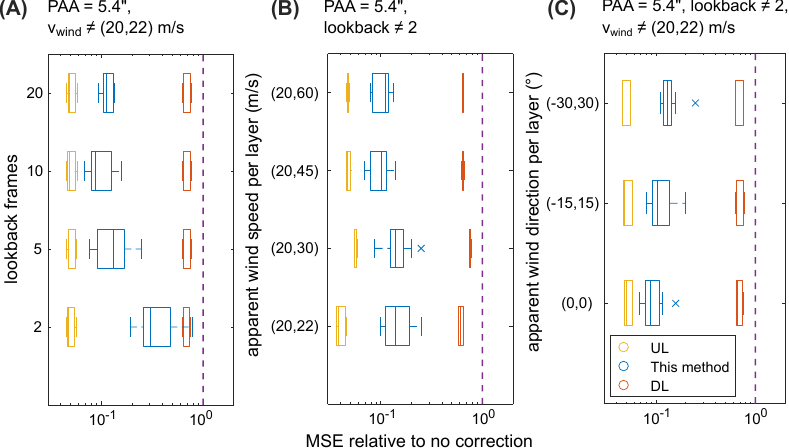}
\centering
\caption[] 
{ \label{fig:mse_vars} 
Box-whisker plots for the uplink phase reconstruction \gls{mse} relative to the uncorrected uplink phase, grouped by (A) the number of lookback frames, (B) the apparent layer wind speeds in the form ($v_1$,$v_2$), and (C) the apparent layer wind directions in the form ($\theta_1$,$\theta_2$). The dashed purple line at \gls{mse} = 1 corresponds to the uncorrected uplink phase reference. Reconstructions for the three cases are shown: UL (yellow), this method (blue), and DL (orange). Specific additional constraints are listed above each plot.}
\end{figure}

The reconstruction errors can also be grouped by the key simulation variables to gain a better understanding of their effect on the reconstruction accuracy. Fig.\;\ref{fig:mse_vars} groups the reconstruction errors by lookback frames, apparent layers wind speeds, and apparent layer wind directions for \gls{paa} = 5.4". Each variable has an intentionally challenging value to highlight where this method struggles. So as not to obscure the trends, a variable's challenging value only appears in its respective subplot. Given the strong correlation between \gls{mse} and \gls{iqr} observable in Fig.\;\ref{fig:mse_iqr}, only the \gls{mse} values will be subsequently analysed for brevity.

Observing Fig.\;\ref{fig:mse_vars}A, the number of lookback frames has the most pronounced effect on the reconstruction error. An increasing number of lookback frames not only greatly improves reconstruction error, but also makes the performance across simulation configurations more consistent. With 20 lookback frames, there is only a factor of 1.5 difference between all configurations at \gls{paa} = 5.4", similar to the DL and UL cases. Interestingly, even at two lookback frames, the reconstruction is nearly always better than the DL case. The practical limit on the useable number of lookback frames is given by the framerate of the \gls{wfs} to remain within the $\sim 20$ ms frozen flow hypothesis timescale. For the 1 kHz framerate simulated, we see that good reconstruction accuracy is certainly achievable within this timescale.

From Fig.\;\ref{fig:mse_vars}B, \gls{mse} trends with layer wind speeds are also clear. The larger the difference in wind speeds, the lower the reconstruction error. In the context of \gls{fso} links with \gls{leo} satellites, the identified equivalent layers will have apparent wind speeds approximately proportional to their heights. This will very likely ensure a diverse and non-degenerate set of apparent wind velocities, which is beneficial to this method. Further, in general, the higher, faster layers will have a lower associated $\mathrm{C_n^2}$, and therefore have less contribution to the reconstruction. For the simulated wind speeds, there does not seem to be a decrease in performance with increasing individual layer wind speeds. Of course, if the wind speed shifts a layer by more than one downlink aperture within a single frame, there would be no correlation between measurement frames, and the reconstruction would fail. This upper limit is given by $v_{\mathrm{layer}}/f_{\mathrm{WFS}} \geq D$, permitting wind speeds of $\sim700$ m/s for the given \gls{ogs} configuration. For a \gls{leo} satellite orbiting at 1\textdegree/s, the corresponding altitude of this equivalent layer would be at 40.1 km and therefore have negligible turbulence associated with it \cite{tysonPrinciplesAdaptiveOptics2022}.

Note that there is a variability in the UL/DL reconstructions with changing layer wind speeds, which should not be the case for a sufficiently sampled atmosphere with a defined Fried parameter. For similar layer wind speeds, the simulation must evolve for longer to capture a statistically diverse atmosphere with the expected \gls{mse}. Differences of $\pm 1$ relative \gls{mse} can be observed for layer speeds of (20, 22) and (20, 30) m/s, respectively.

Finally, from Fig.\;\ref{fig:mse_vars}C, the reconstruction error increases as the apparent layer wind directions stray from the \gls{paa} direction. Of the three variables analysed, this variable has the smallest impact, with a $\pm 30$\textdegree\,difference resulting in a $1.49\times$ increase in median \gls{mse}. Again, due to the dominance of satellite tracking on the apparent layer wind velocities, it is unlikely that the apparent layer wind directions will ever stray this far from the \gls{paa} direction \cite{wangCharacterizingDaytimeWind2019}, except for at the ground layer. To quantify this, let us use the Bufton wind profile. Assume we are tracking a \gls{leo} satellite with a 1 \textdegree/s slew rate, and that the natural wind profile is perpendicular to the \gls{paa} direction. In this case, the apparent wind speed only strays 30\textdegree\, from the \gls{paa} direction below a 600\,m altitude. Even the tropopause at 9.4 km altitude is within $\mathrm{\pm 12}$\textdegree\, of the \gls{paa} direction.

\subsection{Robustness to error}

\begin{figure}[ht]
\includegraphics{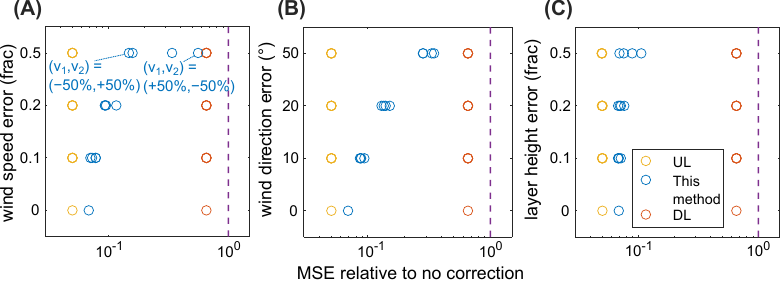}
\centering
\caption[] 
{ \label{fig:mse_errors} 
Scatter plots of the uplink phase reconstruction \gls{mse} relative to the uncorrected uplink phase when error is present in the input parameters. (A) error in the layer apparent wind speeds, (B) error in the apparent layer wind directions, and (C) error in the layer heights. Points are grouped by the absolute value of the fractional errors in Table \ref{tab:error_params}. The dashed purple line at relative \gls{mse} = 1 corresponds to the uncorrected uplink phase reference. Reconstructions for the three cases are shown: UL (yellow), this method (blue), and DL (orange).}
\end{figure}

We can also analyse the robustness of the uplink path reconstruction when error is present in the method's input parameters. A representative configuration is chosen with 10 lookback frames, a \gls{paa} of 5.4" (\gls{paa}/$\theta_0$ = 5), $v_\mathrm{wind,apparent} = (20,45)\;\mathrm{m/s}$, and $\theta_\mathrm{wind,apparent} = (0,0)$\textdegree. Error in the layer heights, wind speeds, and wind directions are then injected according to Table \ref{tab:error_params}. All sign combinations of each error pair are simulated individually, resulting in Fig.\;\ref{fig:mse_errors}.

\begin{table}[ht!]
    \centering
    \caption{Injected error in this method's input parameters. All sign combinations of each error pair are simulated, i.e., 4 combinations per pair.}
    \begin{tabular}{|c|c|c|}
        \hline
        Parameter (symbol) & Error amount & Units\\
        \hhline{|=|=|=|}
        
        \makecell{Equivalent layer\\heights ($h_1$,\,$h_2$)} & 
        \multirow{3}{5em}{\makecell{$(0,0)$\\$(\pm10,\pm10)$\\$(\pm20,\pm20)$\\$(\pm50,\pm50)$}} & \% \\
        \cline{1-1}\cline{3-3}
        
        \makecell{Equivalent layer apparent\\wind speeds $(v_1,\,v_2)$} & & \% \\ 
        \cline{1-1}\cline{3-3}
        
        \makecell{Equivalent layer apparent\\wind directions $(\theta_1,\,\theta_2)$} & & \textdegree \\
        
        \hline
    \end{tabular}
    
    \label{tab:error_params}
\end{table}

Fig.\;\ref{fig:mse_errors}A shows that this method can tolerate wind speed errors of up to $\pm20\%$, with reconstruction error increasing by a factor of $1.25$. Beyond this point, the reconstruction accuracy significantly degrades. Interestingly, the impact of wind speed error on the reconstruction accuracy has a sign dependence, most visible at fractional errors of 50\%. For $(v_1,\,v_2)$ errors of $(+50\%,-50\%)$, the assumed layer wind speeds are much more similar than for errors of $(-50\%,+50\%)$. Vastly different reconstruction errors are observed for these cases due to the performance dependence on layer wind speed similarity established in Fig.\;\ref{fig:mse_vars}B.

The dependence of wind direction error on the reconstruction accuracy is highlighted in Fig.\;\ref{fig:mse_errors}B. A $\pm10$\textdegree\, error results in a modest $1.28\times$ increase in \gls{mse}. Even with errors of up to $\pm50$\textdegree, this method still outperforms the DL case. As previously discussed, when tracking a \gls{leo} satellite, the equivalent layer wind directions will very likely be within $\pm10$\textdegree\, of the \gls{paa} direction, meaning any error in wind direction determination will have a minor impact on the reconstructions.

Finally, we find that errors in the equivalent layer heights have an almost negligible impact on the reconstruction accuracy, with errors of $\pm50\%$ a maximum \gls{mse} increase of $2.0\times$. This is a crucial result, since determination of equivalent layer heights within the atmosphere typically has large uncertainties \cite{hartAtmosphericTomographyArtificial2016}.

Across all three of these key input parameters, the robustness against wind speed error is arguably the most important factor for the performance of this method in a real atmosphere. These simulations currently focus on decomposing an N-layer atmosphere onto N equivalent layers with discrete velocities. However, in practice, the continuous distribution of wind velocity with altitude will lead to a velocity gradient within each equivalent layer. It will become apparent with future modelling whether the identified robustness to wind speed error will be sufficient to handle this velocity distribution over the bounds of each equivalent layer. This, combined with the $\mathrm{C_n^2}$ distribution, will inform the number of equivalent layers required for satisfactory performance of this method.

\section{Conclusion}

We have described a novel method that uses successive measurements of the downlink beam through a telescope's downlink \gls{ao} system to make a tomographic measurement of the atmosphere at several discrete layers. The method takes advantage of the predictable apparent wind velocity profile resulting from satellite tracking to achieve this. By knowing the volume of atmosphere above our telescope, we can reconstruct an estimate of the phase along the uplink path to use as a pre-compensation signal. We find that this method considerably outperforms using the downlink signal directly for pre-compensation, where the improvement scales with the isoplanatic angle to point-ahead angle ratio ($\mathrm{PAA/\theta_0}$). Quantitatively, this method reduces the mean-square phase error and interquartile range of the reconstructed uplink path by factors of up to 8.6 and 15 times, respectively, for an extreme $\mathrm{PAA/\theta_0}$ of 10.

The method assumed frozen flow, and as such is restricted to work within tens of milliseconds timescales \cite{schockMethodQuantitativeInvestigation2000}. We find that, for a standard downlink \gls{ao} system with a 1 kHz \gls{wfs}, this method's performance is not significantly hindered with the 5--20 measurements this constraint permits. We also find that the method is robust to error in its input parameters, namely the assumed equivalent layer heights and apparent wind vectors. Height errors of $\mathrm{\pm 50 \%}$, apparent wind speed errors of $\mathrm{\pm 20 \%}$, and apparent wind direction errors of $\mathrm{\pm 20}$\textdegree\, can be tolerated before the reconstruction significantly degrades.


This method uses the downlink beam/beacon to produce uplink path estimates consistently closer to the ideal case with a beacon at the \gls{paa}, requiring no more constraints on the downlink \gls{ao} hardware than what would already be required for the downlink communications link. This offers impressive hardware simplicity for the tomographic measurement. The actual pre-compensation step would require an additional deformable mirror, and the ability to adjust the launch position of the uplink beam based on the \gls{paa} direction. The latter would be necessary to maximise the overlap between the uplink and downlink paths and could be accomplished using a astronomical derotator. 

Overall, these results are very promising, suggesting that the method could greatly improve the power statistics of uplink \gls{fso} communications outside of the isoplanatic angle -- of particular value for communications with \gls{leo} satellites. Future end-to-end simulations will be conducted to more rigorously validate this method and its impact on signal power at the satellite receiver.

\begin{backmatter}
\bmsection{Funding} Government of Western Australia; Australian Space Agency; University of Western Australia.

\bmsection{Acknowledgments} A.F is supported by Australian Government Research Training Program Scholarships and
top-up scholarships funded by the Government of Western Australia. This work was supported by the
Australian Space Agency’s Moon to Mars Demonstrator Mission program with additional funding from the
Government of Western Australia and the University of Western Australia. This work was made possible through the University of Western Australia Research Collaboration Awards.

\bmsection{Disclosures} The Authors declare no conflicts of interest.

\bmsection{Data Availability Statement} The data generated during this study are available from the corresponding author on reasonable request.


\end{backmatter}


\bibliography{references}

\end{document}